\documentclass{elsart}
\usepackage{harvard}
\usepackage{graphicx}
\usepackage{lscape} 
\usepackage{alltt}  
\usepackage{amssymb,amsmath}
\newcommand{\bk}{\boldsymbol{k}}
\newcommand{\bx}{\boldsymbol{x}} 
\newcommand{\bforce}{\boldsymbol{f}}

\begin{document}
\begin{frontmatter}

\title{Towards Optimal Parallel PM N-body Codes: PMFAST}

\author[CITA]{Hugh Merz}
\ead{merz@cita.utoronto.ca}
\author[CITA]{Ue-Li Pen}
\ead{pen@cita.utoronto.ca}
\author[ASTRO]{Hy Trac}
\ead{trac@cita.utoronto.ca}

\address[CITA]{Canadian Institute for Theoretical Astrophysics, University of
Toronto, M5S 3H8, Canada} 
\address[ASTRO]{Department of Astronomy and Astrophysics, University of Toronto, M5S
3H8, Canada}

\begin{abstract}
We present a new parallel PM N-body code named PMFAST that is freely available to the public.  PMFAST is based on a two-level mesh gravity solver where the gravitational forces are separated into long and short range components.  The decomposition scheme minimizes communication costs and allows tolerance for slow networks.  The code approaches optimality in several dimensions.  The force computations are local and exploit highly optimized vendor FFT libraries.  It features minimal memory overhead, with the particle positions and velocities being the main cost.   The code features support for distributed and shared memory parallelization through the use of MPI and OpenMP respectively. 
 
The current release version uses two grid levels on a slab decomposition, with periodic boundary conditions for cosmological applications.  Open boundary conditions could be added with little computational overhead.  We present timing information and results from a recent cosmological production run of the code using a $3712^3$ mesh with $6.4\times10^9$ particles.  PMFAST is cost-effective, memory-efficient, and is publicly available.
\end{abstract}

\begin{keyword}
Methods: numerical \sep Cosmology: theory \sep Large-scale structure of universe
\PACS 02.60.-Cb \sep 95.75.Pq \sep 98.80-k
\end{keyword}

\end{frontmatter}

\section{Introduction}

N-body simulations are a key tool in astrophysics.  Applications
range from cosmological problems involving dark matter to stellar systems
and dynamics of galaxies.  In many astrophysical problems, $N$ can
be very large.  For precision calibration of statistical weak lensing,
large dynamic range is required and this provides a challenge to existing
computational resources.

A recent development has been the move towards large massively
parallel computers with cheap commodity components and relatively slow
interconnects.  The burden of coding in the presence of a large
memory hierarchy (commonly several layers of cache, local memory,
remote memory, and secondary storage), and distributed message
passing libraries is now placed on the scientist who wishes to
utilize the large machines.

Our goal is to provide the community with a generic N-body code which
runs close to optimally on inexpensive clusters.  
In this paper we describe the design and implementation of the algorithm, and
performance numbers for cosmological applications.

Most real world applications on modern microprocessors achieve a
small fraction of theoretical peak speed, often only a few percent.
An order of magnitude in speedup is available through the use of
assembly coded libraries.  These include routines such as FFT's that
have been optimized to take advantage of the particular benefits
that a given hardware manufacturer can offer in terms of instruction
set and processor architecture developments.

A second limiting factor is the amount of physical memory.  Most
N-body codes are not very efficient in memory use.  In principle, one
only requires 6 numbers per particle to store the positions and
velocities.  In practice, other data structures such as density
fields and force fields dominate memory usage.  

In this paper we present an algorithm that approaches minimal
memory overhead, using only seven numbers per particle, plus
temporary storage which is small.  The computation is off-loaded onto
highly optimized FFT's, and the communication cost on parallel
machines is mitigated by a two level mesh hierarchy.

\section{Optimal Parallel Particle-Mesh N-body}

In this section we describe the physical decomposition of our
algorithm.  Gravity is a long range force, and every particle
interacts pairwise with every other particle.  The use of a mesh
\cite{he88} allows a reduction in computational cost
from  $O(N^2)$ to  $O(N\log N)$.  Unfortunately, FFT's
are highly non-local, and would in principle require global transposes
that move large amounts of data between processors.  This can be costly in terms of network resources, especially in economical parallel clusters 
that employ long latency slow ethernet.

A two level mesh can circumvent this drastic demand on communication
hardware resources.  We follow the lines of Hydra
\cite{couch91}, which decomposes the gravitational force
into long and short range components.  Several authors have described
parallel implementations of particle mesh algorithms.  TPM
\cite{xu95} and GOTPM \cite{dub04}
merge particle mesh and tree algorithms, and have full implementations
of the particle-mesh (PM) algorithm if one turns off the trees.  These
codes are not publicly available, and they were not designed as
optimal PM codes.  \cite{fer95} have also implemented a
distributed memory PM scheme, but which requires significant
bandwidth.

The long range components can be computed on a coarse mesh.  We use a
global coarse mesh which is four times coarser in each dimension than
the fine mesh, resulting in a 64-fold savings in global mesh
communications.  The fine mesh does not need to be globally stored all
the time, as we only need to store the tiles that are being worked
on.  For coarse mesh Fourier transforms, we used the freely available
parallel FFTW library \cite{fftw98}.  This library is based on slab
decomposition, which our current code adheres to.

In order to obtain optimal performance on shared memory multiprocessor
nodes within a clustered environment, the fine mesh is computed on
independent cubic sections of the slab. 
This allows for multiple processors to update fine mesh forces in parallel
and reduces memory overhead by only requiring a fraction of the mesh and its
associated structures to exist in memory at a given time. Coarse mesh
calculations and particle indexing are also parallelized through shared
memory at the loop level to maintain high processor load.  Thread-level parallelization has thus been implemented through the use of OpenMP on the majority of the code, with the only exception being the particle passing routine.  Due to the lack of freely available thread-safe MPI implementations we have limited message passing to only single thread executed regions of the code, a design choice that maximizes portability.

\subsection{Spherically symmetric matching}
\label{sec:sph}

First, we describe the two-level mesh gravity solver as covered in \cite{trac04}. Their method is based on a spherically symmetric decomposition of the potential and force laws.  Here, we consider the decomposition of the gravitational potential, although this method is applied equally well to the direct gravitational force. 

The gravitational potential $\phi(\bx)$ is obtained through a convolution,
\begin{equation}
\phi(\bx)=\int\rho(\bx')w(\bx-\bx')d^3x',
\end{equation}
of the density field $\rho(\bx)$ with a kernel $w(r)=-G/r$.  In order to solve this on a two-level mesh we separate the kernel into a short-range component
\begin{equation}
w_s(r)=
\begin{cases}
w(r)-\alpha(r) & \text{if $r\leq r_c$},\\
0 & \text{otherwise},
\end{cases}
\label{eqn:ws}
\end{equation}
and a long-range component
\begin{equation}
w_l(r)=
\begin{cases}
\alpha(r)\ \ \ \ \ \ \ \ \ \ & \text{if $r\leq r_c$},\\
w(r) & \text{otherwise},
\end{cases}
\label{eqn:wl}
\end{equation}
where the short-range cutoff $r_c$ is a free parameter that will dictate the size of the buffer used between fine mesh tiles and consequently the amount of particles required for passing between nodes.  The function $\alpha(r)$
is chosen to be a polynomial,
\begin{equation}
\alpha(r)=G(a+br^2+cr^4),
\end{equation}
whose coefficients,
\begin{equation}
\begin{aligned}
a&=-\frac{27}{16r_c},\\
b&=\frac{7}{8r_c^3},\\
c&=-\frac{3}{16r_c^5},
\end{aligned}
\end{equation}
are determined from the conditions
\begin{equation}
\label{eqn:kernelconditions}
\begin{aligned}
\alpha(r_c)&=w(r_c),\\
\alpha^\prime(r_c)&=w^\prime(r_c),\\
\alpha^{\prime\prime}(r_c)&=w^{\prime\prime}(r_c).
\end{aligned}
\end{equation}
These restrictions ensure that the long-range kernel smoothly turns over near the cutoff and that the short-range term smoothly goes to zero at the cutoff.

The long-range potential $\phi_l^c(\bx)$ is computed by performing the convolution
over the coarse-grained global density field $\rho^c(\bx)$.  The superscript  
$c$ denotes that the discrete fields are constructed on a coarse grid. Mass 
assignment onto the coarse grid is accomplished using the 
cloud-in-cell (CIC) interpolation scheme with cloud shape being the
 same as a coarse cell.  The long-range force field $\bforce_l^c(\bx)$ is 
obtained by finite differencing the long-range potential and force interpolation  is
carried out using the same CIC scheme to ensure no fictitious self-force.

\begin{figure}[h]
\centering
\includegraphics[width=4in]{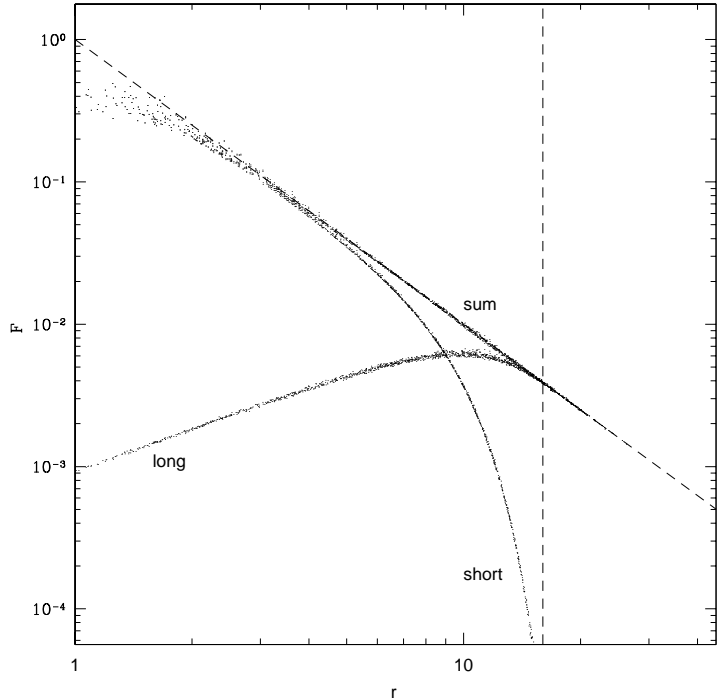}
  \caption{Short and long range force as determined from random particle pairs.  r-axis is measured in fine grid cells, with the short-range cutoff indicated by the dashed vertical line.}
  \label{force_res}
\end{figure}

Since the two-level mesh scheme uses grids at different resolutions, the decomposition given by equations (\ref{eqn:ws}) and (\ref{eqn:wl}) needs to be modified.  In Fourier space, we can write the long-range potential as
\begin{equation}
\tilde{\phi}_l(\bk)=\tilde{\rho}^c(\bk)\tilde{w}_l^c(\bk)=[\tilde{\rho}(\bk)\tilde{s}_\rho(\bk)][\tilde{w}_l(\bk)\tilde{s}_w(\bk)],
\end{equation}
where $\tilde{s}_\rho(\bk)$ and $\tilde{s}_w(\bk)$ are the Fourier transforms of
 the mass smoothing window $s_\rho(\bx)$ and kernel sampling window $s_w(\bx)$,  
respectively.  The mass smoothing window takes into account the CIC mass 
assignment scheme for constructing the coarse density field.  The kernel sampling window
corrects for the fact that the long-range kernel given by equation (\ref{eqn:wl})
is sampled on a coarse grid.  In Fourier space, the corrected short-range potential
kernel is now given by
\begin{equation}
\label{eqn:wscorrected}
\tilde{w}_s(\bk)=\tilde{w}(\bk)-\tilde{w}_l(\bk)\tilde{s}_\rho(\bk)\tilde{s}_w(\bk),
\end{equation}
and can be slightly anisotropic, particularly near the short-range cutoff.
In Figure ~\ref{force_res} we display the contribution to the short and long range force by randomly placed particle pairs on the mesh using the spherically symmetric force matching method.  The errors associated with this data set are shown in Figure ~\ref{sph_force_comp_err}.

\subsection{Least squares matching}
\label{sec:lsm}

In this section we present a
general procedure that we use to minimize the error from the two level
mesh.  The basic strategy is to minimize the error variance in the
total force.  Since a variance is a quadratic quantity in the linear
sum of two kernels, minimization is a linear problem in the value of
the kernel at each point.  We will formulate the problem, and show its
solution.  This generalizes the standard procedure of matching
spherically symmetric kernels as described in Section \ref{sec:sph}.
Since our fine grid is cubical and not spherical, we can utilize its
anisotropy to minimize the force matching error.  We also discuss some
residual freedom in the error weights.

We define the error variance $\epsilon$ of the true force to the grid force
\begin{equation}
\epsilon=\sum_i \left[F^{\rm grid}(\Delta x_i)-F^{\rm exact}(\Delta
  x_i)\right]^2 w_i.
\label{eqn:err}
\end{equation}
Each error term is given a weight $w_i$, which may be chosen to give
constant fractional error, constant radial error, or any other
prescription.
Our goal is to find a grid force law which minimizes the error given
by equation (\ref{eqn:err}).  For a two level grid, we decompose the grid force
into two parts,
\begin{equation}
F^{\rm grid}=F^{\rm coarse}+F^{\rm fine}
\end{equation}
where $F^{\rm fine}$ is given as the numerical gradient of a
potential.   The forces are CIC interpolated 
from the nearest grid cell.
So at each fine grid cell, one has a unique linear coarse force,
and the force is also defined at arbitrary separations.
The primary source of grid error arises
from the inhomogeneity of the CIC interpolation: the force
between particles depends not only on their separation, but also on
their position relative to the grid cells.  Intuitively, one expects
the force error to be minimized when the coarse force is smoothly
varying, since its inhomogeneity is greatest.

In the potential and force calculation, we perform a convolution
over the density field.  To restrict the communication overhead, we
require the fine grid force to be short range, in our case 16 fine grid
cells.  The total number of non-redundant entries in the force kernel
is $n_{\rm fine}=16(16+1)(16+2)/6=816$.  Since equation (\ref{eqn:err}) is
quadratic in both the short range potential and the long range force,
the exact solution is given by the solution of a linear equation.  We
evaluate the sum in expression equation (\ref{eqn:err}) by placing particles at
all integral fine grid cell separations. The
minimization may not be unique, so we use an eigenvalue decomposition
and discard zero eigenvectors.

We note that the pair weighting in equation (\ref{eqn:err}) gives more weight to
large separations since there are more wide separation pairs.  Also,
as written it minimizes the total force error, not the fractional
error.  We use a weight function that weighs pairs
depending on their separation.  In our implementation,
each pair is weighted by the actual Euclidean separation, which corresponds to
constant fractional error.  The coarse grid force at a separation of
zero and one coarse grid cell are also set to be zero.

In the actual implementation, we generate a vector of 816 variables
for the $16^3$ non-redundant entries of the fine grid potential $\phi$,
and a vector of 360 variables to represent the non-redundant three
components of the coarse grid force on a $6^3$ grid.  Call this vector
of 1176 unknowns $\vec{u}$.  We then produce a list of 12320
equations, which over-constrains the solutions.  For each fine grid
cell, we have two sets of three equations, one for each of the three
force components.  We generate equations on an extended $20^3$ grid of
fine grid cells, zero padding the fine grid entries beyond the cutoff.

This results in a set of equations ${\bf A} \vec{x}=\vec{y}$.  The
least squares solution yields $\vec{x}=({\bf A^t A})^{-1} {\bf A}^t
\vec{y}$.  The square matrix ${\bf A^t A}$ may not always be
invertible, so we perform a singular value decomposed solution.  The
actual condition number of the system is $\sim 8.6\times 10^7$ (apart
from singular values).  Double precision is useful to see the
spectrum, where one sees a clear break of eight orders of magnitude
between the zero eigenvalues and the non-zero ones.  Despite the large
amount of over-determinacy, there are 255 singular values which are
left undetermined, and set to zero.  Most of them probably correspond
to coarse grid entries that are at too large separations to be
constrained.  The actual solution took less than one minute on a
laptop.  In contrast, storing the full grid of $32^3$ kernel entries
(which allows one to shortcut symmetries) would result in 64 times
more unknowns, and require a supercomputer to solve the $2^{18}$ times
more expensive problem.

Since we probe only one eighth of one octant in the force kernel, we
need to explicitly enforce boundary conditions on the minimization.
This is done by requiring the long range force to have zero transverse
force along the axes, and to be symmetric along the diagonals.

This optimal force matching results in an anisotropic short range
kernel with cubic support.  This differs from most approaches which
usually impose spherical symmetry on the decomposition.  The resulting errors,
shown in Figure ~\ref{lsq_force_comp_err}, have a smaller scatter than those in Figure ~\ref{sph_force_comp_err}.

\subsection {Algorithm}

The basic logic of the 2-level particle mesh algorithm is presented in Figure ~\ref{algor}.  In this method, particles local to each node are stored in a non-ordered list.  To reduce time spent organizing the particles based on their locations within the mesh, a linked list is constructed by threads in parallel that associates particles contained within each cell of the coarse mesh.  This is achieved by storing the tail of each threads chain as well as the head, allowing for a merger of the individual lists.  Since the linked list is used for determining which particles are to be passed to adjacent nodes it is generated at the beginning of the program execution, as well as following particle passing each time-step.

\begin{figure}[h]
\centering
\begin{alltt}
{\sf 
subroutine particle\_mesh(code)
  if (first\_step) call link\_list
  call position\_update
  call particle\_pass
  call link\_list
  !$omp parallel
  do y\_cube=1,number\_nodes
    do x\_cube=1,number\_nodes
      call fine\_mesh(x\_cube,y\_cube,thread)
    end do
  end do
  !$omp end parallel
  call coarse\_mesh
  call particle\_deletion
end subroutine particle\_mesh
}
\end{alltt}
\caption{Fortran code overview of the 2-level particle mesh algorithm.}
\label{algor}
\end{figure}

Density attribution and velocity updating is implemented with the CIC interpolation scheme on both mesh levels.   On the fine mesh we use a potential kernel and calculate the force by finite-differencing the potential, while on 
the coarse mesh we directly calculate the force utilizing a force kernel.  Direct calculation of the force requires $4$ extra Fourier transforms per coarse time-step, however, this prevents the loss of accuracy associated with finite-differencing and only incurs a minor overhead relative to the fine mesh calculations since there is $64\times$ less data to process.  The code can support any type of
force kernel that one would like to construct and is easily interchangeable.  Kernels generated with the methods illustrated in Sections ~\ref{sec:sph} and ~\ref{sec:lsm} are included with the code. 

\begin{figure}[h]
\centering
\includegraphics[width=4in]{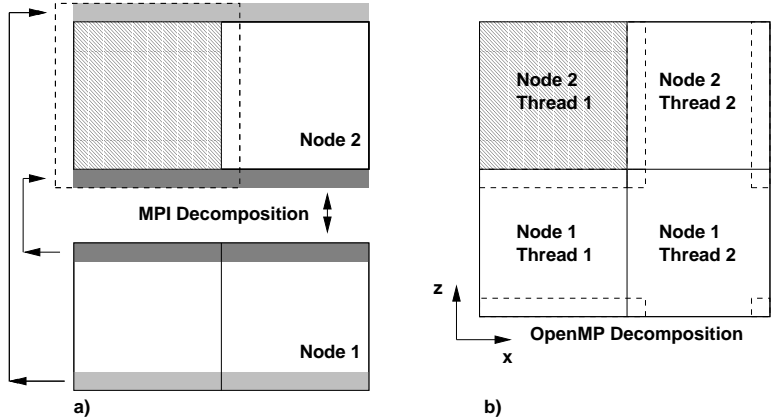}
  \caption{a) An example of the data decomposition on the fine mesh
  for 2 nodes using 2 threads per node.  The local data for each fine mesh is bounded by the
  solid line and the mesh boundary including buffer region is the dashed
  line. The buffer regions acquired from the adjacent node are
  indicated. b) The same data set, showing the fine mesh overlap.  The
  fine grid is only stored for the region that is actively worked on,
  which reduces memory overhead.}
  \label{decomp}
\end{figure}

Each fine mesh cube requires a buffer density region along its surface area to
calculate fine mesh forces within the fine range force cut-off.
For the dimensions perpendicular to the decomposition this can be readily
obtained using the particles in the slab, however, additional information is required about
the density in the dimension along the decomposition from adjacent nodes.  This
layout is presented in Figure ~\ref{decomp}.

We communicate buffer particles from adjacent nodes to locally calculate the
density in the buffer region. This approach removes the communication dependency
from mesh calculations and allows it to be done in tandem with the passing of migratory particles, minimizing local processing cost as well as removing potentially complicated communication patterns that could lead to deadlock.

The slab decomposition of the physical volume guides the approach that is
used to pass particles between nodes.  Referencing of particles based on their
location within the mesh is implemented through a linked list spanning the
coarse mesh.  The interface between nodes is searched using the linked
list to determine if particles lie within the region for buffer construction or
migration. Particles that migrate are indexed in an additional deletion list
and all of the particles to be passed are included in a buffer for passing.  The
passing then occurs over all nodes synchronously and is repeated in the other
direction, re-using the buffer.  This is a suitable approach for cosmological
applications as the particle flux is relatively balanced between nodes and is 
dominated by the buffer region.   

\begin{figure}[h]
\centering
\includegraphics[width=4in]{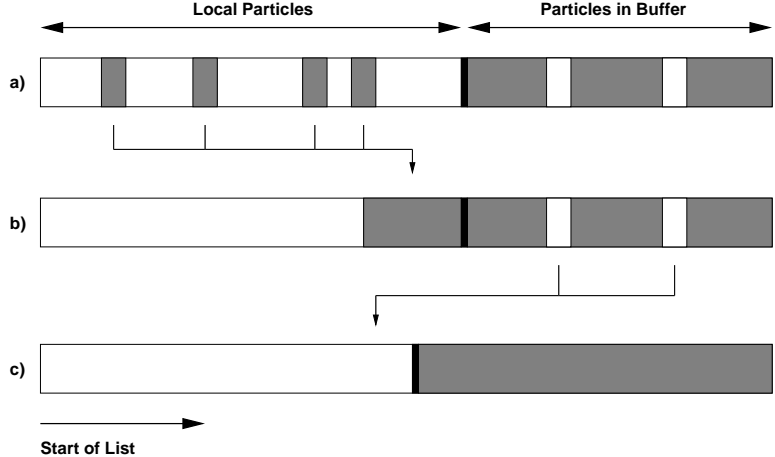}
  \caption{Sorting of the local particle list during particle passing.  a) list following passing of particle buffers b) list following sorting of particles that are migrating out of node c) list following sorting of particles that have migrated in.  Dark regions represent particles that are to be deleted from the list at the end of the current time-step.}
  \label{partlist}
\end{figure}

Rather than an additional loop at the end of the step for deletion of buffer
particles and particles that exited the node, this process is done during the
passing routine and is illustrated in Figure ~\ref{partlist}.  Particles that are to 
be deleted are shuffled to the end of the particle list using the deletion list.  
The incoming buffer region is then searched for new local particles and these 
are swapped to the end of the now contiguous local particle list.  In this fashion 
one need only change the index of the total number of particles in the list at 
the end of the step to delete particles that lie outside of the local mesh.

In an effort to maintain a high processor load we have developed a file transfer interface for the MPI FFTW library. The MPI FFTW routines currently are not thread-safe and rather than having only one thread per node participate in the calculation of the coarse mesh Fourier transform we execute a separate program which runs an FFTW process for each processor on each node.  While this offers no gain in performance for single processor nodes it can provide nearly linear speed-up on multi-processors with a memory overhead equal to that used to store the coarse grid density.  The data that is to be transformed is first decomposed on each node by the PMFAST process into a number of slabs equal to the number of processors that the node contains.  This data is then written to a file-system, at which point the FFTW processes read in the data, perform the transform and write it back.  The PMFAST process then reads the decomposed slab and resumes operation.  By using temporary file-systems in RAM we avoid the latency cost of having to communicate this information to disk.   

\subsection{Multi-stepping}

We use the time step constraint 
\begin{equation}
\Delta t =\sqrt{\Delta x/a}
\end{equation}
where $\Delta x$ is the grid spacing and $a$ is the maximal gravitational
acceleration.  The coarse grid has four times the grid spacing, and a
smoother gravitational field, so the time step is typically limited by
the fine grid.  We can exploit this and compute the coarse grid forces
less frequently than the fine grid.  Second order accuracy in time can be
maintained using Strang-type operator splitting.

The code currently supports a variable time-step scheme in which
multiple fine grid updates are performed for every coarse grid step.
This is currently done in an $N:1$ ratio, where $N$ is an odd integer
and represents the number of fine steps calculated per coarse step
sweep.  In order to maintain second order accuracy all of the
fine-steps within the sweep are calculated with the same time
interval, and the coarse step is done halfway through the sweep with
$N\times$ the time interval used for the fine steps.  The length of
the time-step is variable and limited by the maximum acceleration
calculated on both the fine and coarse meshes as well as expansion to
maintain integration accuracy.  For example, if $N=5$, we perform two
fine step, one combined fine-coarse step, followed by two more fine
steps.  The code computes and displays the maximal acceleration on the
fine and coarse grids at each time-step.

We have already described a range of design choices which minimizes
memory and network requirements.  To further optimize the code in the
presence of memory hierarchies (cache), we use the linked lists in
each coarse mesh cell to compute densities and update velocities.  On
the fine grid, the forces are computed by taking the gradient of the
potential on a small sub-grid.  Then we loop over all particles on the
fine sub-grid, which leaves the forces in cache.

\subsection{Boundary conditions}

For cosmological applications, we use periodic boundary conditions.
The advantage of the two level mesh is that isolated boundary
conditions are easily applied.  The standard procedure of using a
kernel of twice the size of the computational domain usually results
in an eightfold computational cost penalty.  In the two level grid, we
only need to double the coarse grid.  Even a doubled coarse grid is
only 1/8th the size of the fine grid, and still a small cost for the
whole computation.  This is currently not implemented in the code.

Cosmological initial conditions for use in simulations with PMFAST can
be obtained from the website for the code, or one may employ another
generator such as grafic2 \cite{bert01}, a Gaussian
random field generator which can be obtained from
http://arcturus.mit.edu/grafic/

\subsection{Accuracy}

Our goal is to be able to control errors to enable precise cosmological
simulations with a goal of achieving 1\% accuracy on the non-linear dark matter
power spectrum down to scales below one Mpc.  This is about an order of
magnitude smaller than the non-linear scale.

\begin{figure}[h]
\centering
\includegraphics[width=4in]{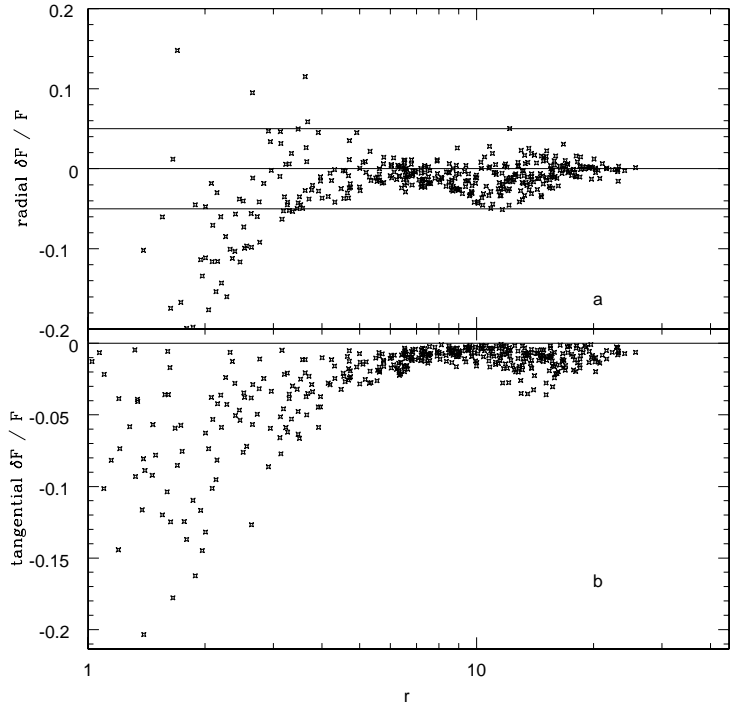}
  \caption{Fractional error in force resolution generated with random pairs using the spherically symmetric matched kernel: a) fractional radial force error b) fractional tangential force error.  The r-axis is measured in fine grid cells.}
  \label{sph_force_comp_err}
\end{figure}

\begin{figure}[h]
\centering
\includegraphics[width=4in]{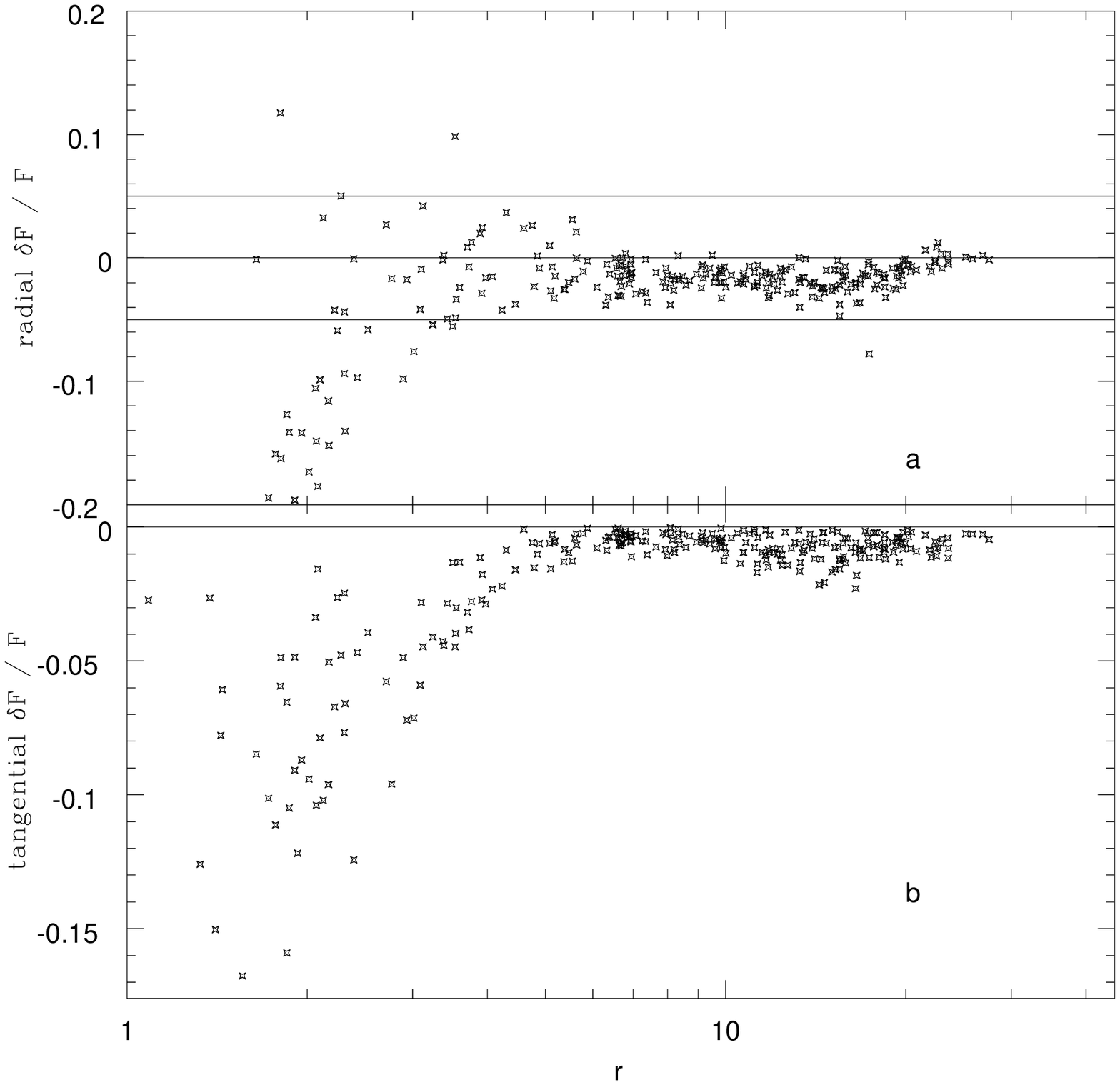}
  \caption{Fractional error in force resolution generated with random pairs using the least squares method matched kernel: a) fractional radial force error b) fractional tangential force error. The r-axis is measured in fine grid cells.}
  \label{lsq_force_comp_err}
\end{figure}

Errors arise from a range of approximations.  The grid forces deviate
at the grid scale, the coarse-fine overlap scale, and on the box
scale.  Particle discreteness leads to Poisson noise.  And the finite
time step leads to time truncation error. The contribution of the fractional error for randomly placed particle pairs in both the radial and tangential directions is displayed in Figure ~\ref{sph_force_comp_err} using the spherically matched kernel and ~\ref{lsq_force_comp_err} using the least squares method matched kernel.  

\begin{figure}[h]
\centering
\includegraphics[width=4in]{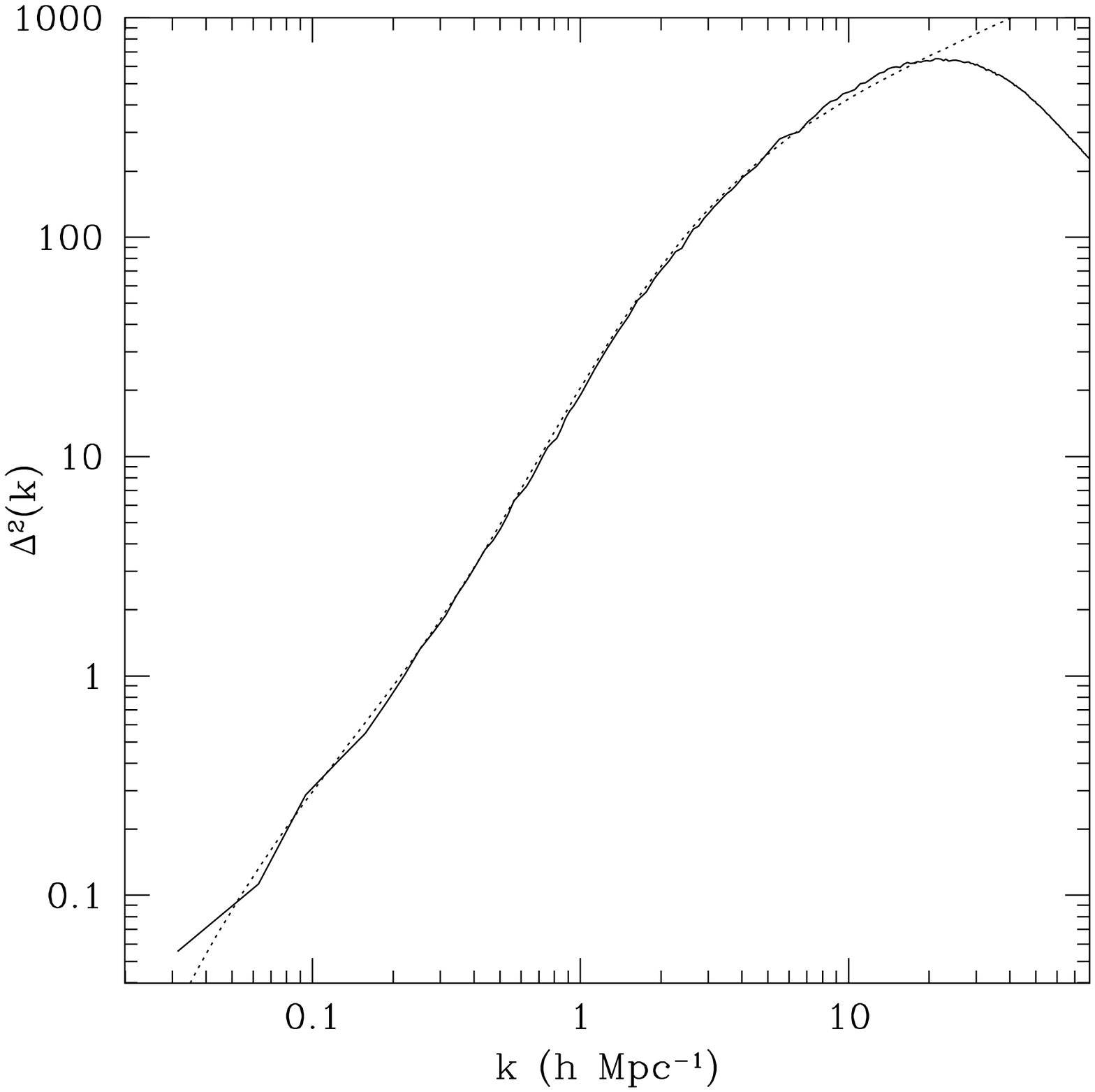}
   \caption{Dark matter power spectrum comparison between a $3712^3$ cell mesh simulation using 6.4 billion particles (solid line) and the spectrum as computed using the halofit algorithm at a redshift of 0 (dotted line)} 
   \label{pow_spectrum}
\end{figure}

In order to gauge the cosmological accuracy of the code we have included in Figure ~\ref{pow_spectrum} a comparison between a $3712^3$ mesh simulation computed using PMFAST and the power spectrum as generated by the halofit algorithm \cite{sp03}
.  Generation of the spectrum from the simulation data requires more memory
than is currently available in any single node.  We thus plotted a spliced curve composed of three spectra calculated from
the same distribution.  Inspection of the power at different wavebands was 
achieved by first scaling the data-set to a $1024^3$ mesh, followed by 
scalings to $4096^3$ and $16384^3$ grids which were then folded into
64 and 4096 cubes respectively and superimposed onto a $1024^3$ mesh.

\begin{figure}[h]
\centering
\includegraphics[width=4in]{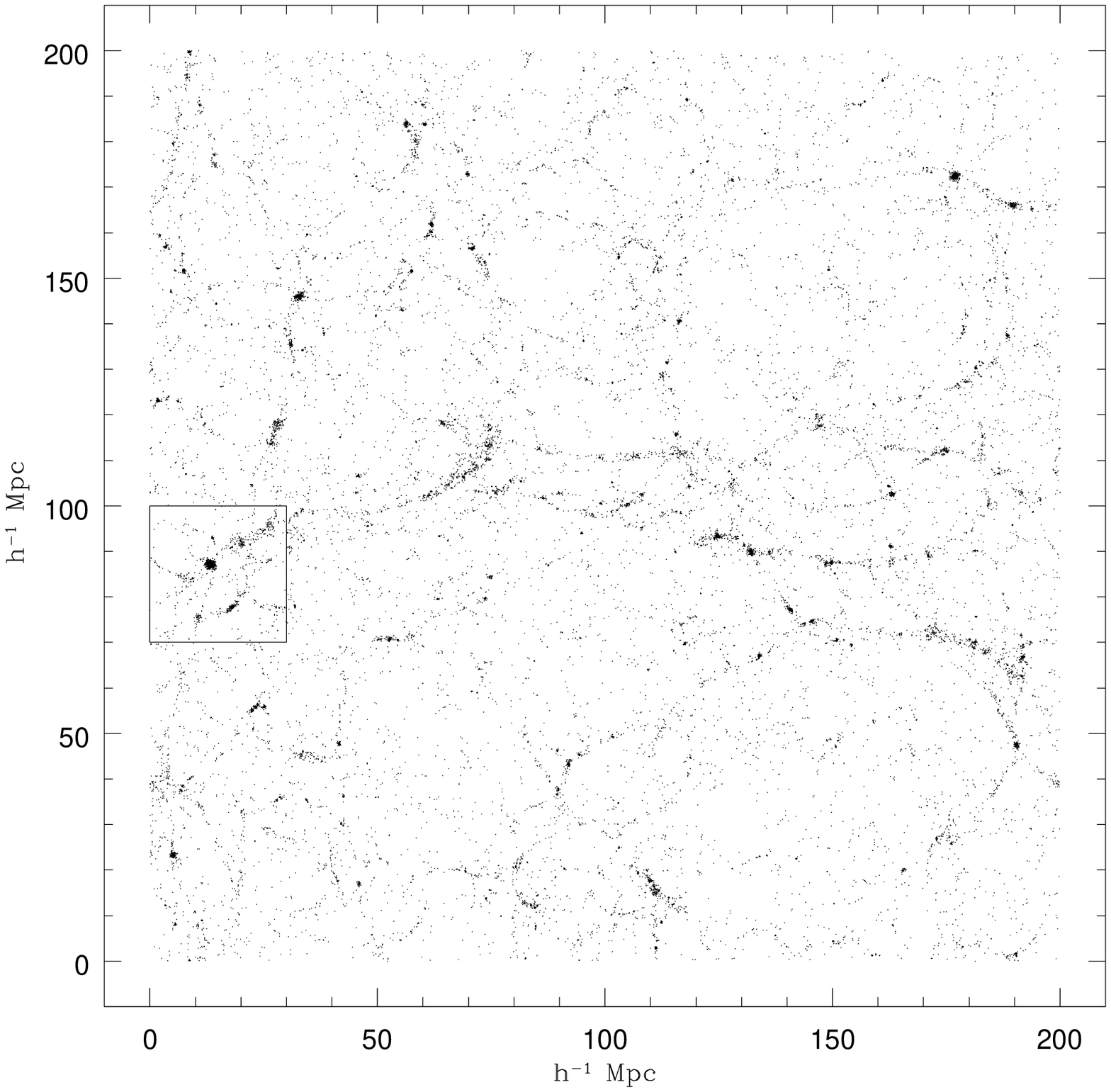}
  \caption{10 kpc particle projection slice taken from a $6.4\times10^9$ billion particle cosmological simulation at a redshift of 0.  The inset is shown in Figure ~\ref{fig:slice_inset}.}
  \label{fig:slice}
\end{figure}

\begin{figure}[h]
\centering
\includegraphics[width=4in]{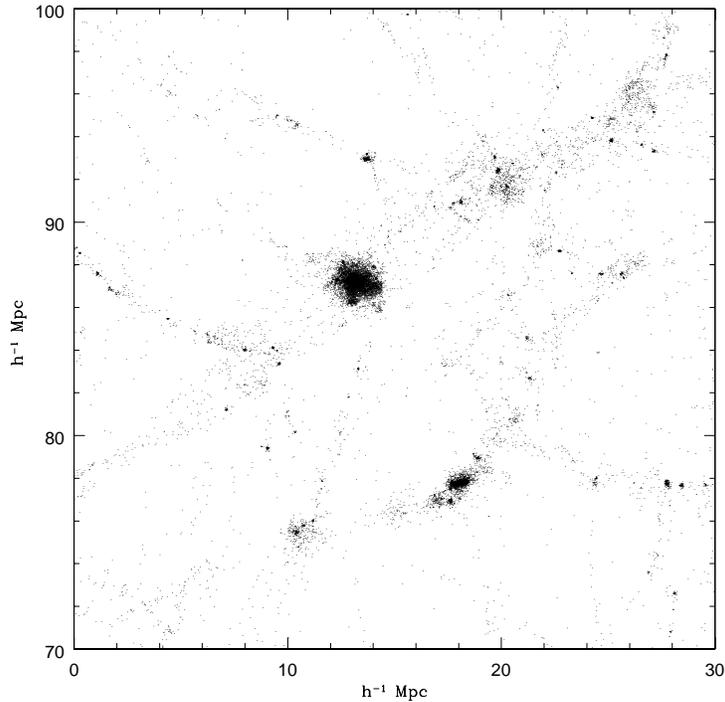}
  \caption{Inset from the particle projection displaying the region of densest clustering in higher resolution}
  \label{fig:slice_inset}
\end{figure}

\begin{figure}[h]
\centering
\includegraphics[width=6in]{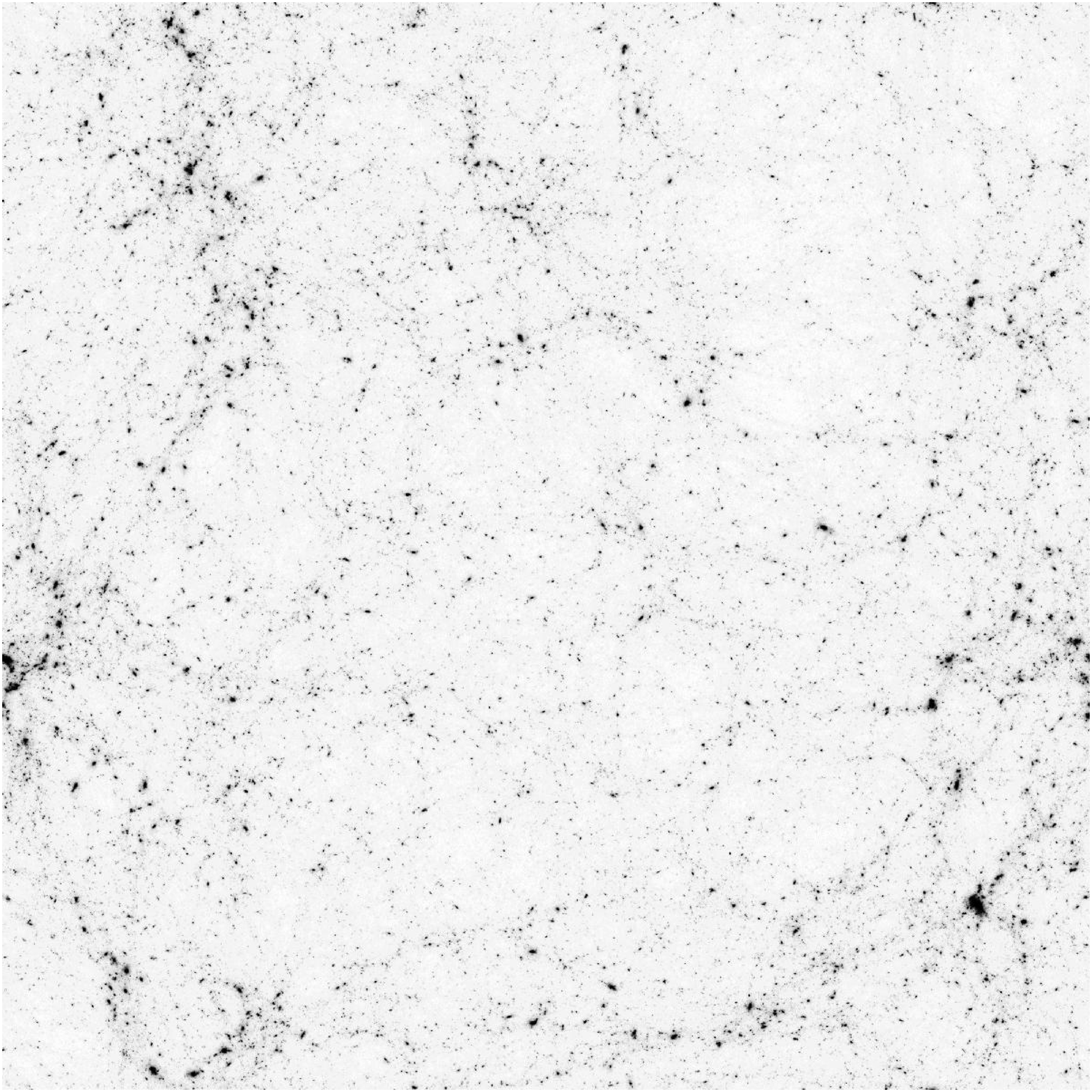}
  \caption{$3712^2$ cell density projection at a redshift of 0.033 calculated from the $6.4\times10^9$ particle simulation. The box width is 200 $h^{-1}$ Mpc.}
  \label{fig:rho_proj}
\end{figure}

Figure \ref{fig:slice} shows the distribution of particles within a $10$ kpc
thick slice.  A region is shown in higher resolution in Figure
\ref{fig:slice_inset}.  The code also generates on the fly
two-dimensional density projections, which are used for weak
gravitational lensing analysis.  The projection of the density field
to the mid-plane is shown in Figure \ref{fig:rho_proj}.

\subsection{Timings}

Our production platform is an IA-64 cluster consisting of 8 nodes,
each of which contains four 733 MHz Itanium-1 processors and 64 GB RAM.  The cluster has point-to-point gigabit ethernet connections between each node. The maximum fine mesh grid size that we
have run is $3712^3$ using $6.4\times10^9$ particles.  The total fine mesh grid length depends on the number of nodes used in the simulation, the width of each fine mesh cube (dashed boundary in Figure ~\ref{decomp} and the buffer length such that
\begin{equation}
\rm{Grid} = (\rm{Cube} - 2 \times \rm{Buffer}) \times \rm{Nodes}
\end{equation}

With moderate clustering and a maximum particle imbalance of 12\% from
mean density each time-step sweep at a 5:1 fine to coarse ratio takes
approximately $2100 s$ to complete.  This time does not vary
significantly for minor load imbalances.   The time estimate is
obtained by taking the time taken for 5 fine steps plus one coarse
step, and dividing by 5.  The fine grid FFT's account for less than
20\% of the computation time.

\begin{figure}[h]
\centering
\includegraphics[width=4in]{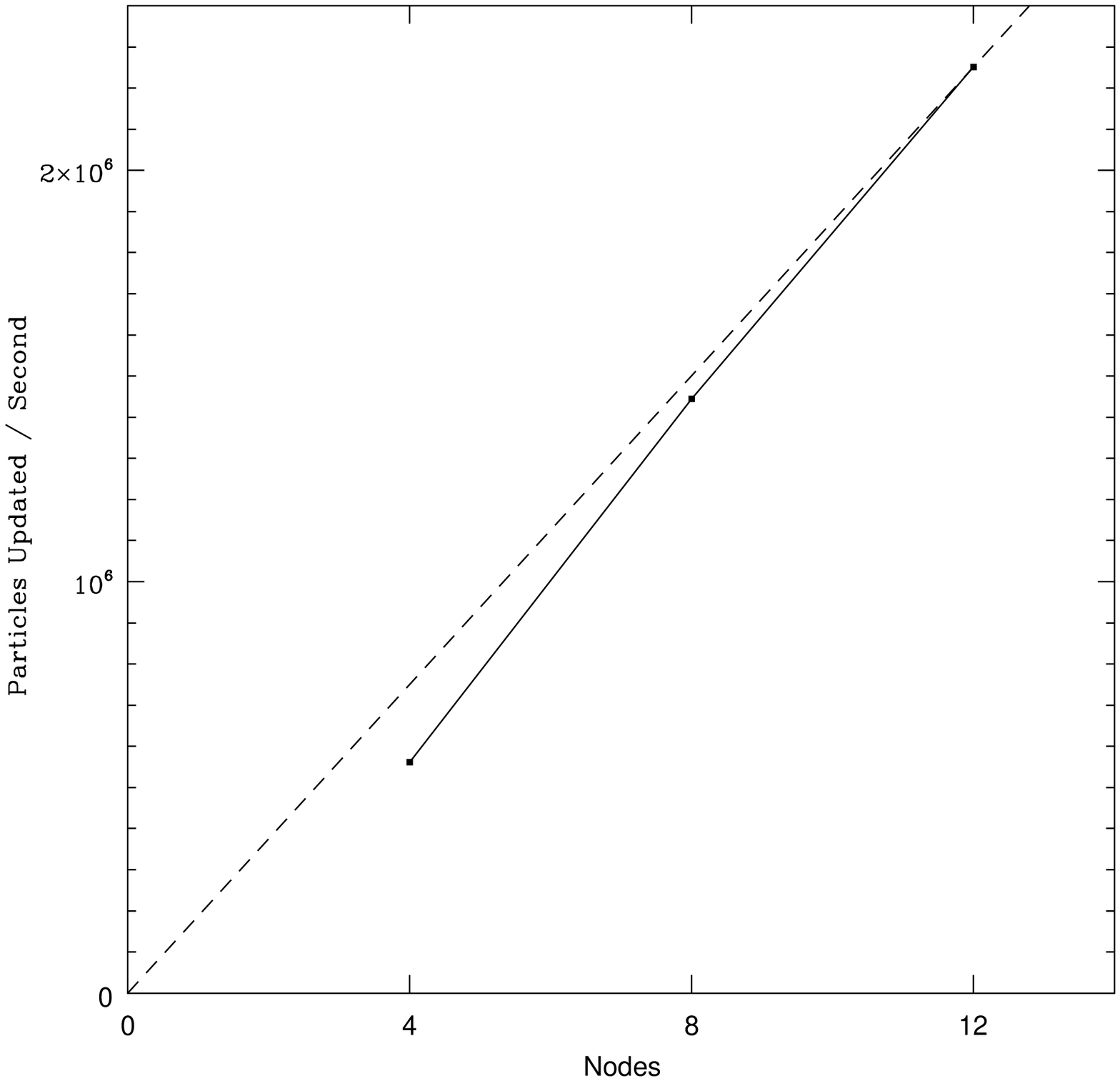}
  \caption{Weak scaling profile as measured on the IA-32 platform.  Linear scaling based on the 12 node rate is indicated by the dashed line.}
  \label{wscale}
\end{figure}

The code has also been timed on the local CITA Beowulf cluster
\cite{dub03}, composed of dual Xeon 2.4 GHz processors
nodes with 1 GB ram and gigabit ethernet using smaller grid
sizes. Table ~\ref{timing} includes timing data for the simulation on
the two platforms using a $5:1$ fine to coarse mesh time-step ratio.
We performed a weak scaling test, where the size of each fine sub-grid
is 128 grid cells.  This results in an effective 80 usable fine grid
cells after overlap.  In this regime, the overlap makes density
assignment and fine grid FFT's a factor of 4 inefficient.  On the IA-64
production platform this overlap only accounts for 25\% of the volume using
a fine mesh of 512 cells.
Due to the overlap between fine grids, it is not easy to time a pure
strong scaling test while keeping the total grid size fixed.  Our fine
grid was restricted to be a power of two. The code also exhibits good
performance under weak scaling on the IA-32 platform, becoming memory
limited at 12 nodes utilizing a $960^3$ total fine mesh.  The weak scaling 
curve is displayed in Figure ~\ref{wscale}.

\begin{table}
\caption{Timing Results on IA-32 and IA-64 Platforms}
\centering
\begin {tabular} {l r r r | r}
\hline\hline
Platform& \multicolumn{3}{c}{IA-32}& IA-64 \\
\hline
Nodes&4&8&12&8 \\
Particles/Node&$1.0\times10^6$&$4.1\times10^6$&$9.2\times10^6$&$8.0\times10^8$ \\
\hline
Position Update&0.1&0.3&0.7&99.6 \\
Particle Passing&0.8&3.6&7.9&262.1 \\
Link List&0.2&0.9&2.1&60.1 \\
Fine Mesh&3.6&14.7&34.8&1,513.8 \\
Coarse Mesh&2.7&3.2&3.6&166.2 \\
\hline
Timestep&7.3&22.7&49.1&2,101.8 \\
Particles/Sec&$5.6\times10^5$&$1.4\times10^6$&$2.3\times10^6$&$3.0\times10^6$ \\
\hline
\end{tabular}
\label{timing}
\end{table}

\section{Future Expansions}

\subsection{Cubic decomposition}

The current code works on a one-dimensional slab decomposition, which limits the
degree of parallelism that can be achieved before surface area
effects begin to dominate the computing cost.  The local CITA
Beowulf cluster has 256 nodes, which makes a 1-D decomposition across 
the whole cluster impractical.  Of course a 3-D
decomposition can be implemented along the same lines, which is in progress.

\subsection{Multi-level}
\label{sec:multi}

The current code works on two levels.  This dictates the number of
overlap cells needed between coarse and fine grid forces.  In
principle, one could use a larger number of grids, and reduce the
overlap range by a factor of two.  Similarly, one can trade-off the
global communications bandwidth with the local buffer size.  In a
multi-level implementation, only the top level would be done globally.
This could be on an even coarser grid than our current implementation.
If one passed density fields instead of particles, the buffer regions
would also be hierarchical.  The communication costs are then
dominated by the overlap between the finest and second finest grids,
which could be reduced to 8 fine grid cells.  The buffers for the
coarser cells are still 8 grid cells on each coarsened level, but
these are a factor of 4 cheaper, and asymptotically only add up to a
1/3 overhead.  The total coarse grid must be at least a factor of eight
finer than the width of the logical computer lattice if one does not
want buffers to span more than the nearest neighbors.  For the
proposed Universe Simulator with 10000 nodes, this would be 21
processors on a side, corresponding to a $168^3$ coarse grid, whose global
communication is completely negligible.

\subsection{Out-of-core}

With the current speedup and efficiency, the code is memory limited on
most existing machines.  While we have already reduced the memory
overhead close to the theoretical minimum, one could gain many orders
of magnitude in capacity by implementing an out-of-core design in which
simulation data is cached to disk \cite{trac04}.  In such a scheme, a multi-level grid as described in the previous section would be needed.

\section{Conclusions}

We have presented a new freely available parallel particle-mesh N-body
code that takes a significant step towards achieving optimality in CPU,
communication and memory performance.  The only $O(N)$ memory required is
six floating point and one integer per particle.  A two level force
decomposition allows for the use of a short range force which minimizes
communication.  It also eliminates the need to store a global fine
grid density field.  CPU performance is optimized by the use of vendor
optimized FFT libraries, which allows one to deploy very fine grids. The
code is available for download at:

http://www.cita.utoronto.ca/webpages/code/pmfast/

\end{document}